# CHAPTER ONE

## 1.1 INTRODUCTION

In an evolving Nigerian banking industry, strategies are being adopted by the major players in order to achieve their long-term organizational goals- profitability and survival. In the light of this belief, much emphasis is being laid on the computerization of their banking operations. Within the last decade, the Nigerian banking industry has been at the forefront of computerization. This is with the aim of: improving their information system, delivery of efficient and high quality service to their customers.

In the advent of computerization, old generation banks in the industry were force to wake up from their slumbers and face reality. The new generation banks came into the industry with innovation, research and development, in order to push these less competent old ones out of business. This revolution was aided by computerization. It's very pertinent to state that computerization is a very powerful weapon which can be employed to annihilate competitors. The Nigerian banking industry ensures that their data base is updated as at when due. With the 2004 banking reform in Nigeria, the industry has persistently remained vibrant. The overall aim of this sanitations effort by Professor Charles Chukwuma Soludo was to achieve some macroeconomic objectives,which include the following :

a. To create a world of banking groups through mergers and acquisitions aimed at avoiding bank distress.

b. To consolidate Nigerian banks to manage the country's external reserve deposited in foreign banks

c. To ensure that there is zero-tolerance in the regulatory framework.

According to Rob Kling (1996), "when a specialist discusses computerization and work, they often appeal to a strong implicit image about the transformation of work in the last one hundred years and the role that technology has played in some of these changes." In view of this, it's quite imperative to analyze this information with the Nigerian banking industry, and to take actions aimed at guiding against the problems associated with computerization.

## 1.2 STATEMENT OF THE PROBLEM

The twenty-first century business-world has been marked by the rise of service job in such areas as: banking, insurance, advertising, transportation and health. Many of the earliest commercial computers were bought by large service organizations such as: banks and insurance companies.

One of the basic problems of this research work is to critically look at the banking industry vis-à-vis the heavy investment made by these institutions in the acquisition of computer technology. This will be geared towards assessing the effect of this computer technology on the banking activities in general.

With computerization, jobs that would have taken hours and a lot of manpower, can conveniently be done by few hands in just a jiffy. The concomitant effect of this will inevitably be lay-off of workers. In the case study of this research work- United Bank for Africa Plc- about 4000 workers were made redundant as a result of technological innovation. According to Shankar K. (1997) "as society becomes more dependent on computers, computer crime is becoming not only more disastrous, but also more attractive to the criminally-minded people: the disclosure threat, which is the dissemination of information to an individual for whom that information shouldn't be seen.

In the context of computer security, the disclosure threat occurs whenever some secret information that is stored in a computer is divulged to somebody who shouldn't have seen it.

Fraudulent people have devised a clever way of guessing at people's passwords. These could take the following forms: a user's role model, surname, best movie, spouse's name, street name, best drink.

Another major problem associated with computerization is obsolescence. Technology, which is one of the external threats in the business environment, changes very rapidly. What was impossible yesterday has been made possible today. With technological changes, huge investments made by banks in the acquisition of contemporary computer devices are lost. This has a very high cost implication on the resources of these banks.

## 1.3 OBJECTIVE OF THE STUDY

The objective of this study is to carry out a thorough analysis of effect of computerization on the Nigerian banking Industry. The positive side of it will be fully look at, and the problems associated with it will be brought out in details. This study is also aimed at taking a holistic view of computerized banking operations in order to help identify clearly those beneficial aspects of computerization on the Nigerian banking industry. Advantages of computerization such as: Speed, efficiency, reliability and accuracy will be thoroughly analyzed.

The proper management of the computerization process aimed at cushioning its traumatic effect like job losses will also be delved into. The use of computers in the Nigerian banking Industry has been necessitated by the large volume of book-keeping and their multifarious scale of operations. In studying the effect of computerization in the Nigerian banking industry, efforts will be made towards analyzing issues such as:

a. Computer fraud

b. Job-losses

c. Obsolescence of computer systems.

This research is also aimed at providing an in-depth analysis of computerized banking operations which will serve as a guide for any bank that intends to computerize its operations.

## 1.4  RESEARCH QUESTIONS

Data analyzed should be able to answer the following questions that will be used in appraising the effect of computerization in the Nigerian banking industry:

a. Has computerization increased service delivery in the Nigerian banking industry?

b. Does computerization escalate the incidence of fraud in the Nigerian banking industry?

c. Can job-losses be the end-product of computerization in the Nigerian banking industry

d. Can computerization bring about less fatigue in the Nigerian banking industry?

e. Can computerization affect the internal control system in the banking industry?

f. Can data be secure in the computer systems?

## 1.5 RESEARCH HYPOTHESIS

$H_0$: Computerization has improved service delivery in United Bank for Africa Plc.

$H_1$: Computerization has not improved service delivery in the United Bank for Africa Plc.

## 1.6 RESEARCH METHODOLOGY

Data for this research will be obtained through the following:

a. **Questionnaire:** this will be administered on the staff of UBA, who are in-charge of information technology. This will involve, close-response questions with only 'yes' or 'No' answers to assist in carrying out the study.

b. **Library Research:** this will be carried out with the aim of consulting relevant literature already written on the subject matter.

c. **Personal interview:** this will be employed to obtain response to question of every technical nature.

## 1.7 SCOPE AND LIMITATION OF THE STUDY

A research on the effect of computerization on the Nigerian banking industry is a very vast one. It involves an extensive study and understanding of the total effect of computerization on the Nigerian banking industry, considering its merits and demerits. In view of the large size and scale of operation, it is a very painstaking and excruciating one. In a bid to keep the work within a manageable limit in terms of time and financial constraints, the research will be limited to United Bank for Africa Plc.

It's quite pertinent to state that banks have a discreet nature of not divulging sensitive information about their scale of operation. And, this has served as a limiting factor. The problem of bias on the part of the respondents might serve as a limiting factor.

Despite these factors, efforts will be made to obtain the much-needed information to carry out this research work and the conclusion drawn to be incisive.

## 1.8 SIGNIFICANCE OF THE STUDY

The significance of the study of the effect of computerization on the Nigerian banking industry could be established from the point of view that one of the important benefits which a computer system could bring to an organization is speed and accuracy of information. The study is further justified by the need to guide the banks in their quest to computerize their operations, thereby bringing out the cost implication of doing so.

The need to take a critical look at the Nigerian banking industry which is the life-blood of the Nigerian economy makes this study very important.

In view of the fact that no economy can make headway without a sound and vibrant financial system makes this study very important.

## 1.9 BACKGROUND OF THE STUDY

According the Oxford English dictioary (6$^{th}$ edition), "computerization is an act of providing computers to do identified operations". Before the advent of computers, the rule of thumb adopted in data-handling led to incompetence and huge loss of resources.

According to Rob Kling(1994), "In the early twentieth century, the organization of office work underwent substantial change-firms began to adopt telephones and typewriters. By 1930s and 1940s, many manufacturers had devised electromechanical machines to help manipulate, sort and tally paper records automatically."

In the early 1960s, business firms such as insurance companies and banks adopted computer-based information systems on a large scale.

During the 1980s and early 1990s, virtually every business organizations had bought the personal computer (PC). The PC revolution did not only change the nature of office work but, also expanded the range of people who used the computer.

In the last decade, many larger organizations that imbibed computerization and saw their profit rise astronomically concluded that "computerization is the currency to be spent in the new millennium".

In the light of this belief, this study probes the pros and cons of computerizng banking operations, using the United Bank For Africa plc as a case study.

## 1.10 HISTORICAL BACKGROUND OF THE UNITED BANK FOR AFRICA PLC

The United Bank for Africa Plc (UBA) has its history dated back to 1949, when the British and French Bank ltd (BFB) commenced business in Nigeria. Following Nigeria's independence from Britain, UBA was incorporated in 1961 to take over the business of BFB, with its headquarters at 127/129 Broad Street, Lagos. The bank's current MD is Phillip Odozua.

Today's United Bank for Africa Plc is the product of merger of the old UBA and the former Standard Trust Bank Plc, and a subsequent acquisition of the erstwhile Continental Trust Bank (CTB). The union emerged as the first successful corporate combination in the history of the Nigerian banking industry.

Since the historical emergence from the merger with former Standard Trust Bank Plc, the UBA Group has positioned itself to be Nigeria's dominant bank and a leading player in the African continent. In the year 2000, Europe's frontline Finance and economy magazine.

Euro money named UBA the best domestic bank in Nigeria, in recognition of its exponential growth in the past couple of years and the comparatively high inflow of investment from global Finance players. And in 2007, Pan-African News Magazine awarded UBA the Emerging Global Bank award, indicative of the international bank which has most positively influenced the African Continent.

The United Bank for Africa Plc has consistently positioned itself as the bank to be reckoned with in the Nigerian banking industry. It has grown its total assets by over 345.01% from ₦198. 68 billion ($1.656 billion) to ₦884.14 billion ($7.368).

UBA has the largest distribution network in Nigeria with over 6.5 million customers. As at 30$^{th}$ September 2008, it had over 656 business offices, 296 deploy Pos and 1332 ATMS. It has staff strength of over 14000. Regionally, the group has its presence in 18 countries under the direction of its GMD? CEO, Mr. Phillip Oduoza, the management team of the group is made up of people with skills in various fields.

# ORGANISATION OF UNITED BANK FOR AFRICA (UBA)

UBA group's operating structure is organized around seven strategic business units (SBUs) and four Strategic Support Units (SSUs), informed by the need to reinforce its leadership in service delivery, relationship management and the execution of its strategy. In addition, the GMD/CEO is supported by the Group Executive Office consisting of the strategy office, the corporate transformation office, the chief of staff and adviser to the GMD/CEO.

## UBA Strategic Business Groups (SBGS)

- ❖ **UBA Plc Nigeria- North**

  UBA Nigeria-North covers the northern region of Nigeria, whose economy is largely agrarian in nature.

  UBA Africa was set up to take advantage of trade financing opportunities across continent.

- ❖ **UBA International**

  UBA international has been created to provide wholesale, correspondent, commercial, customer and transactional banking services outside Nigeria. It has driven the bank's expansion into UK, America, and Asia .

**DEFINITION OF TERMS**

- **Computerization:** This is the act of providing computers to perform identified operation.

- **Reform**: A change that is made to an organization or a social system in order to improve it.

- **Obsolescence**: The state of becoming old-fashioned and no longer modern.

- **Fraud**: A crime of deceiving somebody in order to get money or good illegally.

- **Data Base**: An organized set of data that is stored in a computer and can be accessed and used in various ways.

# CHAPTER TWO

# LITERATURE REVIEW

## 2.0    INTRODUCTION

**Computer and Organization**

Modern-day business organizations strive to achieve certain goals for the benefit of their owners and clients, these goals are usually expressed in terms of objectives such as: increase in turnover, cost reduction, profit maximization and improvements in services. Such objectives will need to be met within the confines of the available resources.

In order to meet these objectives, an organization should be able to:

a. Plan ahead
b. Control cost
c. Coordinate its activities

Today, large business organizations have discovered that the best means to deliver high quality service is by embracing computerization.

## 2.1    WHAT IS A COMPUTER?

According to G.B Davies (1978), a computer can be defined as any electronic device that can accept data, process this data and produce this processed data as output, at a very high speed and accuracy with little or no human intervention.

## 2.2 HISTORICAL EVOLUTION OF COMPUTERS

The existence of computers can be traced to the early days of man. As a matter of fact, the computer has become known to mankind as early as $7^{th}$ century BC when the first computer called the ABACUS came into being. It was made In China and consisted of beads strewn on iron rods. The abacus was used mainly for simple arithmetic calculations such as additions and subtraction. Since the invention of the Abacus, computer has metamorphosed through different stages before its present form.

**The Adding Machine**

According to Omilo Augustine(1996 ),the adding machine was invented in 1962 by a Frenchman called Blaise Pascal, a nineteen- year- old boy, who was motivated by the need to reduce the numerical labour involved in his father's work(the supervision of the tax and expenditure of public money).

In 1823, a breakthrough was recorded in what is today known as computer. That year, a British mathematician called Charles Babbage "the father of computer" invented a machine called the difference Engine, which computed tables rapidly and accurately.

He didn't receive the expected assistance because his ideas continued to develop beyond the original concept. In fact, due to lack of fund, he abandoned the project and commenced work on the second machine called "The Analytical Engine". However, parts of the difference engine were made and it can be found in the science museum in south Kensington, England.

The Analytical Engine was patterned towards the modern-day computer. For instance, it had the processor, input device, output device, main storage and the control unit. The Analytical Engine could be programmed. It held its programs on punched card. Charles Babbage could not complete his work on Analytical Engine before he died in 1872.

The punched card was first used for automatic control of the weaving loom in 1801 by Jacquard. However, it became useful for computational work at the tail-end of the 19$^{th}$ century. At that time, an American statistician by name Herman Hollerith who was working in the US census department suggested its use to the government. He devised a card-based machine which he used for the processing of the US census data in 1890.

Herman Hollerith later left the census office in 1896 to form the tabulating machine company which later became what is today known as international Business machine (IBM) inc. one of the largest computer manufacturers in the world, and a pioneer in the production of micro-computers in commercial quantity.

In 1944, an automatic calculator called the Automatic Sequence Controlled Calculator (ASCC) was invented. This later became known as mark-1. This feat by Howard Aiken of Harvard University was encouraged by the success of the punched card machine in 1937. He was greatly assisted by IBM which provided all the manufacturing capabilities.

The success of the Mark-1 led to the invention n of both Mark-2 and Mark-3. All of which used the punched card as input medium. The output device was a card punch and a typewriter and it operated electro-magnetically.

The Second World War propelled a lot of efforts in the development of electronics. The evidence of this can be found in the design of a machine called the Electronic Numerical Integrator and Calculator (ENIAC).

Then, the best was yet to come as John Von Neumann led other scientists at Advanced studies institute at Princeton, to develop EDVAC-Electronic Discrete Variable Computer. This could be said to have set the stage for the manufacture of today's computers. As a result of Neumann's achievement, a great number of computers were designed in the United States of America, and elsewhere.

## 2.3 CLASSIFICATION OF COMPUTERS

Computers can be classified under two broad categories:

a. By Data processed
b. By Size

### BY DATA PROCESSED

Considering the manner in which data is represented within the computers, we could have following classes.

- **Analog Computers**

These are computers that operate on values represented in the form of continuous variables. That is, values that change frequently within short time interval. In analog computers, a value is represented by a physical quantity that is proportional to it.

The output from this type of computers is represented in the form of smooth curves or graphs from which information can be read. A physical quantity used may be current, voltage, pressures, temperature and length.

- **Digital Computers**

These are computers that process data which is represented in the form of discrete values (e.g. 0, 1, 2, 3) by operating on it in step. Discrete values occur at each step of operation. Counting on one's fingers is probably the simplest digital operation we all know. Today, there are more applications that use digital rather than analog computers.

- **Hybrid Computers**

This is a computer that combines the speed of digital computers and the efficiency of analog computers.

B**. BY SIZE**

- **Mainframe Computers**

This is a set of computer that is large and expensive, which is used by many people for a variety of purpose.

**Features of Mainframe Computers**

a. They contain general-purpose processor capable of handling multiple simultaneous functions
b. They support a wide range of peripheral equipment including high speed storage devices .

c. They are normally housed in air-conditioned rooms, surrounded by security devices.

- **Mini Computers**

These are computers that are smaller in size and have a very low cost of purchase. Mini-computers are easier to install than mainframe computers.

a. no need for complex management structure.

- **Micro Computer**

In physical size, the micro-computer is the smallest. It's also known as "Personal Computer" (PC). It's very cheap to acquire a micro-computer. It's commonly used in homes and small offices. It's portable and can tolerate the normal room or office temperature. Again, its operations are very simple and therefore require little or no skill. It's simple to install, and maintain. However, it's slow and has a limited storage capacity.

**Types of Micro Computers**

i. Palm top

ii. Laptop

iii. Desktop

## 2.4　COMPUTERS GENERATIONS

1.　**The First Generation**

The first generation computers were characterized by the presence of 'vacuum tubes' which acted as a basic block for building the logical part of the computer. It consisted of circuits made up of wires and values. Examples of computers in this generation include UNIVAC1, IBM, 700. The period covered by the first generation computers lasted from the Mid Nineteen Forties to Mid Nineteen Fifties.

2.　**The Second Generation**

This generation of computers was characterized by the use of transistors in place of vacuum tubes. This made the computers consume less electricity power and produces less heat. They were physical smaller and cheaper. The computers in this group include IBM 7030, NCR 315, UNIVAC 1107 and Honeywell 800. The period of the second generation computers covered late fifties to about mid Nineteen Sixties.

3.　**The Third Generation**

In this computer generation, the integrated circuits tech neology was the major characteristics of the computers. This made computers cheaper than the earlier ones and these computers were also faster as they operated in nano seconds (one billionth of a second). This generation witnessed the introduction of many mini and mainframe computers. The major companies involved included: IBM, Honeywell amongst others. These computers were used for more complex data manipulations.

4.  **The Fourth Generation**

This generation witnessed the production of micro –computers. It was characterized by the use of micro-processors; it came along with software such as

   a. DBASE: for datable management.
   b. Lotus 1-2-3: for spreadsheet applications.
   c. Word-star for word processing.

This generation enhanced the use of computers in communication, especially in terms of improved networks and electronic mail. The fourth generation covers mid nineteen seventies to late eighties.

5.  **The Fifth Generation**

Computers in this generation came into the market around 1990. This era witnessed the introduction of micro-computers. These computers are data-driven.

2.5  **CHARACTERISTICS OF COMPUTERS**

1. **Speed:** data are processed at a very high speed.

2. **Accuracy**: data are processed without error.

3. **Consistency**: A computer, if given the same task, it'll continue to produce the same result.

4. **Automatic:** It's automated with minimum human intervention.

## 2.6 FUNDAMENTAL COMPUTER SYSTEMS

The main components of the computers system are:

a. Input Devices.

b. Central Processing Unit.

c. Primary Storage Unit.

d. Secondary Storage Unit.

e. Output Devices.

a. **Input Devices**: Are used to capture and enter data into the computer. Before data can be used within a computer system, it's usually necessary to convert it into a format that supports processing by computer. Most data are held in machine-sensible form. We shall briefly discuss some examples of input devices:

i. **The Keyboard**

This remains the most common input devices and its basic design has remained largely the same.

A common criticism of the keyboard is that inexperienced users find it difficult to use.

ii. **The Mouse**

This is a device held by the user to navigate from one side of the screen to another. Selections like menu items are made by clicking one of the buttons on the mouse. A mouse is suitable for controlling programs that make use of icons, menus or buttons.

b. **Central Processing Unit**

The central processing unit (CPU) performs processing by carrying out instructions given in the form of computer programs. It's made up of two components:

i. The control unit

ii. The arithmetic and logic unit.

The control unit fetches commands from memory, decodes them and then executes them. It controls the operations of all hardware, including all input/output operations. The arithmetic and logical unit carries out arithmetical calculations.

c. **Primary Storage or Main Memory**:

This is the memory that store and load data and instructions temporarily. Such memory is said to be a volatile memory and it is:

i. Random Access Memory(RAM):

   The content of this type of memory are lost when the power to the device is switched off.

d. **Secondary Storage Devices**:

They provide a means of storing data and programs until they are required. Secondary storage devices include the following:

i. **Floppy disk drive:**

   It uses plastic disk, coated with a magnetic covering and enclosed within a rigid plastic base. They come in different sizes. At present, the most common type is the high density 3.5" floppy disk.

ii. **Hard disk drive**

   This is a standard feature of a modern computer. It is used in storing the computer's operating system, application software and data. It stores data on a number of rigid patterns that are rotated at high speed.

iii. **The CD-Rom Drive**

This arose from the audio compact disc player and began to gain popularity in the late '80s. The acronym 'CD-Rom' stands for Compact Disc-Read Only Memory. Data cannot be written to a CD-Rom by a conventional player.

- The digital versatile Disc(DVD), although, this is similar to a CD-Rom, two significant benefits to users are:

d. **Output Devices**

This translates the result of processing into a human-readable form. The results of a calculation, for example, can be displayed on a screen or sent to a printer. Output devices include:

➢ **Video Display Unit (VDU)**

This is the most common output device, also known as the monitor. Here, information can be shown instantly, and the cost of using the monitor as an input device is very low.

➢ **The Printer**

The printer is a very common output device and thus, it's considered an essential part of a computer based management information system. The basic types are: LASER, DOT-MATRIX, INKJET and DAISY WHEEL.

> Laser printer:

This is a type of printer that makes use of the toner powder it's usually described as page printers. It has a high quality printing ability, and it's almost completely silent in operation.

## 2.7 THE COMPUTER SOFTWARE

A software can be define as a set of programs designed to run on the computer so that the computer can perform the function for which its designed. A program is a series of coded instruction showing the logical steps the computer should follow to carry out a function, the software can be classified into two: system software and application software.

a) **System Software:**

This refers to the suite of programs that facilitates the optimal use of the hardware systems and provides a suitable environment for writing, editing, debugging, testing and running of user programs. Usually, every computer comes with a collection of this suite of programs which is provided by the hardware manufacturer.

b) **Application Software:**

This includes programs designed to solve problems of specific nature. It could either be supplied by the computer manufacturer or in some cases, the users produce their own application programs called "USER PROGRAM". Hence, the application software can be sub-divided into two classes:

i. Generalized software.
ii. User-defined software.

**(i) Generalized Software.**

This includes word-processing software such as

- Page-maker
- Ventura
- Corel draw
- Spreadsheet

Stated below are the various types of generalized software's:

❖ **Word-Processing Packages**.

Examples are word-perfect, WordStar, display writer, professional writer, Lotus manuscript, ms-word.

❖ **Spreadsheet Packages**

A spreadsheet is a sheet of paper divided into grids of rows and columns on which you can do financial (numeric) calculations. A spreadsheet calculation program does exactly the same thing on the screen of your computer. Examples are: LOTUS 1-2-3, supercalc, ms-Multiplan, Informix and excel

## 2.8  COMPUTER OPERATING SYSTEM

An operating system (OS) is a suite of program acting as an interface between the users, and it provides the user with features that make it easier for him to code, test, execute, debug and maintain his program by efficiently managing the hardware resources.

One of the suites called the executive program remains resident in the main store and controls all the other programs. Running a user program involves several steps:

i. Initiating of language processor.
ii. Provision of data.
iii. Initiation and execution of programs.
iv. Removal of results.

If a computer system has no operating system, it follows that most of these tasks will be carried out by the operator. In essence, the processor will be idle most of the time, which will affect the output of the system. The output of the system precludes this by reducing the operator's intervention to the barest minimum.

**Types of Operating Systems**

In a computer system, the operating system can be classified basically into three classes:

a. **Single-User System**

This kind of operating system makes the machine available for only one user at a time. Examples are the operating systems found on the micro-computers otherwise known as Personal Computers (PC) which are: MS-DOS, PC-DOS, Windows 95, and Windows 98.

b. **Multi-User System**

This is another class of operating systems that is capable of managing and coordinating the hardware resources of multi-user computer systems. The multi-user is an environment where a computer called the 'host' is shared by many through terminals (i.e. console and key board). Examples are: UNIX, PC-MOS, and XENIX

c. **Network Operating System**

In this class of operating system, computers are fully or semi-indecently connected together with this connection, various resources such as printers, CD-Rom, file server are being shared by the connected systems. Examples of operating systems in a network environment are:

## 2.9 General computer business applications.

Computers are used in virtually all the field of endeavor such as:

a. Engineering
b. Law
c. Medicine
d. Banking
e. Public Administration.

In the area of business, computers have extricated business managers of the pain of the safe-keeping of vital data needed for the continued survival of the entity. Because of its versatility, many business operations can be performed on it.

### 2.9.1. Computer application for information

Business organizations need a suitable and reliable information processing system. This processed information must satisfy the needs of the government, shareholder customers, labour union and other stakeholders in the organization. For instance, the government requires a quarterly or annual report of the income taxes of the citizenry. While state and local government agencies require report of sales taxes collected, in the same veil, the central bank of Nigeria (CBN) requires all banks to file their monthly returns on loans and advances with them (the CBN). Also, a business enterprise must furnish annual reports to stockholders and various information to customers, creditors and the general public. To meet these objectives, computerization is the easy and ready-made weapon to achieve this.

According to L.C Okorie (1987 page 30), operational information processing on the other hand is performed to advise management on the status of the various units of the company, so that appropriate action can be taken. This is the control aspect of business information processing and the need is internal nature. It assists the management in assessing the level of progress made by measuring the performance of personnel.

This, since management needs information from all parts of the organization, one of the major assets of recording, manipulating and report data. The best facility for this purpose is the computer. Some data are important only at certain times and places in the business cycles. For instance, information about accounting, purchasing returns may be needed when a firm is bidding for business. After the contract has been signed or lost, such information may be of little or no value. For a piece of information to be of great value, it should be:

i.  Available when needed.
ii. Available where needed.
iii. At the right level of accuracy.

iv. At the right quality

v. Gathered, processed and reported at a reasonable cost

Another advantages that computerization has helped the business is in the area of payroll preparation. The calculation of wages and salaries involves a number of variables, but common factors which relate to the personal details of each employee. Generally, information generated through the computer system will be invaluable to the line managers in decision-making. The personal director would require up-to-date information to monitor all corporate expenditure so as to ensure that budgets are not unnecessary overshot.

The store controller needs an up-to-date record of inventory. In other words, the decision to carry out an assignment depends on the information available to the manager and the confidence that the manager has in the information. In addition, experience has shown that the procurement of good information that would make for efficiency and economy in modern times has proved too difficult a task for manual data-processing system.

Computers are therefore, the justifiable alternative to an otherwise slow and fraud-ridden manual information processing system. In summary, J.N Forester (1968, page 37) said that "if management is the process of converting information into action, then it's clear that management's success depends primarily on what information that is chosen and how the conversion is executed".

## 2.9.2. Computer Application for Solving Problems.

Computers are used in business organization for solving problems. However, computers do not use the human approach to solving problems. It must be told where to find required information, what logic to use, what calculation to perform and what form of output to produce. However, computers can only process data in machine-readable form. Instruction coded in other language must be translated into a processing language with the aid of a computer or an assembler.

## 2.10. Data flow In Banks

The effective operation of banks is dependent on the efficiently flow of data, both internally and externally. Accurate, prompt and complete data is essential for quick customer services and operation of the various banking activities. Bank managers and officers alike, must have adequate data to properly direct the operations of their banks.

The operation department is usually at the centre of the data-generation and dissemination units of banks. This is why banks computerize their operations to enable operations to be carried out effectively and efficiently.

For computerization to be of optimum value to a bank's operation, certain data must be available at specific point in time in the banking cycle. A system of processing data must take the element of time into account. Because, the efficiency and effectiveness of any good service delivery must be timely.

The storage and retrieval of information is an important function of data-processing. The accuracy and thoroughness of any effective data-processing is measured in terms of quality and timelessness, at the right place.

## 2.11 FACILITIES FOR THE DEVELOPMENT OF A MANAGEMENT INFORMATION SYSTEM (MIS) IN BANKS

There are several facilities needed for the development of an efficient information system in banks. The facilities are generally compliant, and they include:

- Self service terminal
- Point of sale terminal
- On-line network
- Data base

### 2.11.1    SELF SERVICE TERMINAL (SST)

Self service terminal (SST) is an electronic device that facilitates a wide range of banking activities, but doesn't accept cash deposits. They are accessible, easy to use and more cost-effective, and helps one avoid long queues. The following transactions can be performed on an SST.

   a. Make payment to third-party beneficiaries.

   b. Obtain information in foreign exchange rates.

### 2.11.2　　POINT OF SALE TERMINAL (POS)

This is a computerized replacement for the manual cash register. It includes the ability to record and track customers' orders, process debit and credit cards and mange inventory. A pos terminal system for a restaurant, for example, is likely to have all menu items stored in a database that can be relied upon.

### 2.11.3　　ON-LINE NETWORK

This is a system of competently linking workstation in order to eschew the problem of congestion. The on-line network is very advantageous in the following ways:

   a. It enables serial devices without changing software and hard ware.

   b. It enables real-time data analysis.

   c. It enables remote access monitoring and control.

### 2.11.4　　DATABASE

A database is a well-structured repository for data. The overall purpose of such a repository is to maintain data for some set of organizational objectives. Most database systems are built to retain the data required for the running of the day-to-day activities of an organization.

Hence, in a university, a database system will be needed for such activity as recording the continuous assessment of students.

**PROPERTIES OF A GOOD DATABASE**

**(a)  Data Sharing:**

Data stored in a database is not usually held solely for the use of one person. A data base is normally expected to be accessible to more than one authorized persons.

Hence, a student's database might be accessible to members of academic and administrative staff.

**(b)  Data Integration**

Shared data brings numerous advantages to the organization. Such advantages, however, only result if database is treated responsibly. One major responsibility of database usage is to ensure that data is integrated.

**(c)  Data Integrity**

Another responsibility arising as a result of shared data is that a database should display integrity. In other words, a database should accurately reflect the universe of discourse that it is attempting to model.

**(d)    Date Security**

One of the ways of ensuring data security is by restricting access to the database. The main way this is done in contemporary database systems is by defining in some detail a set of authorized users of the database. For instance, a secure system would be one where the finance department has access to information used for collection of students' fees, but is prohibited from changing the fee levels of given students.

**What is a Database Management System?**

A database management system (DSMS) is an organized set of facilities for accessing and maintaining one or more database. A DBMS is a shell which surrounds a database and through which all interactions take place within the database. The interactions catered for falls into 3 main groups:

**(1)    File Maintenance**

- Adding new files to the database
- Removing files from the database
- Updating data in existing files

**(2)    Information Retrieval**

- Extracting data from existing file
- Extracting data for use by application programs

**(3)    Database Control**

- Creating and monitoring users of database
- Restricting access to files in the database.
- Monitoring the performance of database.

**2:11:5     ELECTRONIC FUNDS TRANSFER SYSTEM (EFTS)**

According to Olowe R.A (2002), this is a system of transferring funds from one country to another via an electronic means. This system of funds transfer is only exclusively reserved for commercial banks.

The bank will then debit the customer's account and credit either directly or send the money by bank wire transfer system, to the supplier accompanied by electronic information supporting the transaction. For EFTs to work effectively, the following conditions must be present:

- Both the customer and the company must use the same bank whose branches are integrated electronically.
- If the customer and the company use different banks, both banks must be integrated electronically.

## 2.12 ORGANIZATIONAL CHART OF A COMPUTER DEPARTMENT

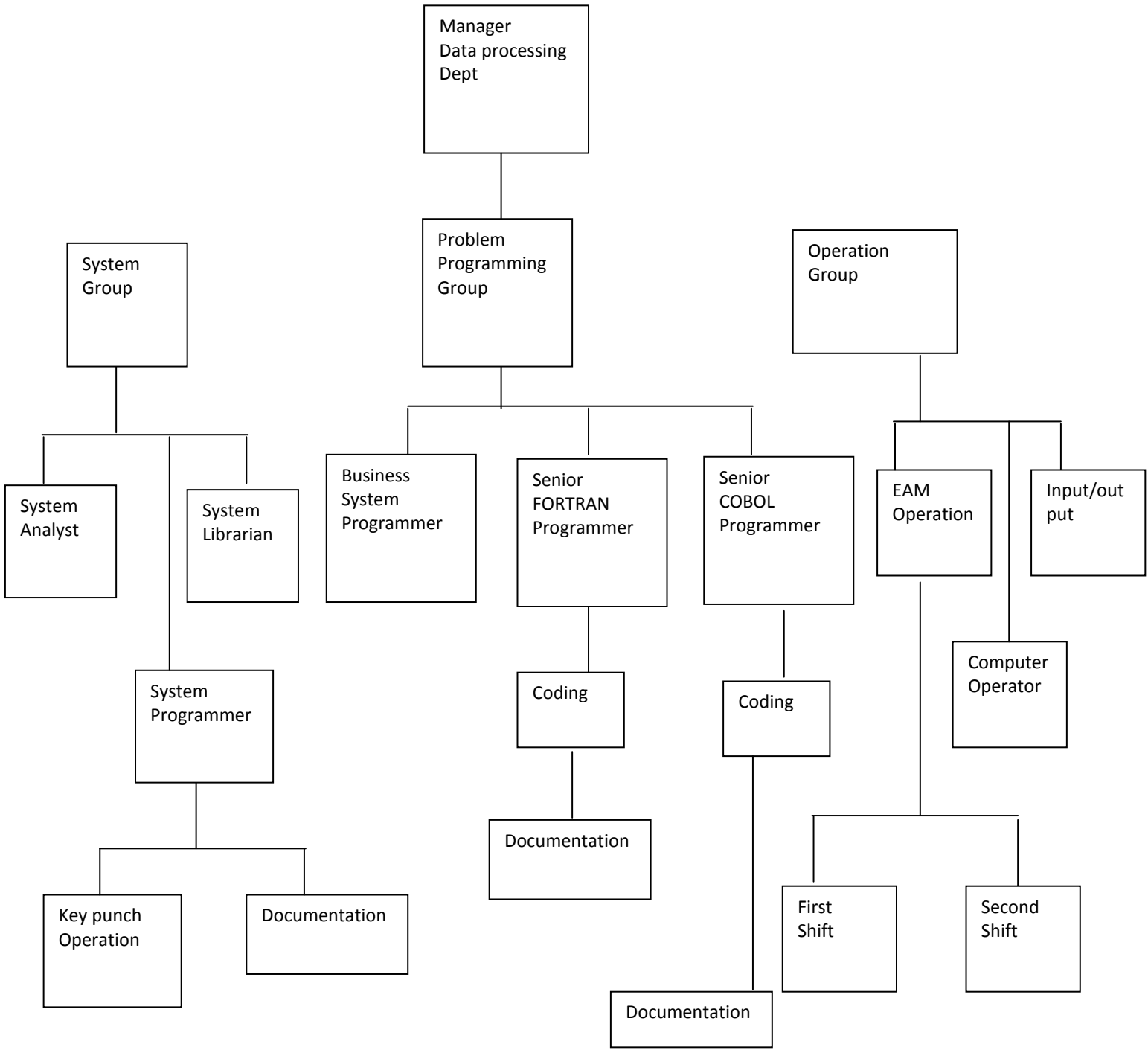

## 2:13  INFORMATION SYSTEM PROCESSING

**(a)  Batch Processing**

With the increase in number of data to be processed, it makes more sense to gather and store individual data (of the same type) in batches, so that when the data have accumulated to a certain number, they are sorted and processed at once. The processing time is predetermined by the designer of the application and depends mainly on the operational role of the batch within the organization. Processing may be periodically i.e. hourly, weekly, monthly, or after a specific number of data is attained.

Batch processing was the earliest mode used in commercial data processing and it is currently used in some situations where the transaction data comes on paper such as processing of cheques and credit card slips.

**Advantage**

1. Efficient and cost-effective use of the computer resources

**Disadvantages**

1. The gap between data capture and information generation renders this mode of processing unacceptable in an organization that requires immediate results or information to influence other operations or decisions.

2. This mode allows for delay in the operational system, because errors detected after batch processing may take some additional time to correct.

### b. Real-time Processing

The expression "real-time" is applied to any system, which produces almost immediate responses as a result of information input. The essential feature of this information processing is that the input information is processed quickly enough so that action can then be promptly taken on the results.

Real-time processing forms the basis for a type of information system called 'Decision Support System'. This is because the system depends upon immediate access to information stored in the computer system. In this mode of information processing, a dual-purpose input and output devices, such as terminals and VDU's are required. Communication front-end is necessary to interface lines connecting remote terminals to the CPU.

**Advantages**

1. Immediate response is given to all inquiries
2. It is often interactive
3. The amount of paper circulated between departments is reduced as information can be displayed on the user terminal.

**Disadvantages**

1. The systems that operate in real-time are often complicated and relatively expensive: initial costs include programming, testing higher skilled personnel, file protection,and back-up equipment.

2. A breakdown in the system may be disastrous if no back-up facility is available.

**C    Centralized Processing**

Centralized processing is a processing mode that is accomplished either by having all the computers in the organization in a centralized computer centre or by giving one large central computer, with telecommunications links to other micro-computers at other locations. This mode of information processing is often more efficient because it tends to eliminate redundant resources and duplication of effort, since all data or information are processed at the central point. department.

**2:14    PREREQUISITE FOR THE DEVELOPMENT OF A COMPUTER-BASED MANAGEMENT INFORMATION SYSTEM (MIS).**

Good planning is a prerequisite of any business activity, and computer data processing is no exception.

A properly planned computer-base management information system would offer significant improvement in management information reporting. All plans should be thorough and would as much as possible be close to reality. The planning for the introduction of computerized management information system in the banks requires all level of planning (strategic, tactical and operational). And should cover all areas of information needs within the bank's top management: head office admin, regional or area administration.

**A.    Strategic Planning**

This phase would come about as a manifestation of policy decision of the top management of the bank. The policy should define the long-term objective of the bank, and strategic planning would form part of the overall corporate plan.

The strategic plan would help keep the banks as a going concern and would span over several years usually between 5 years and 10 years. This phase would consider units and the entire contribution of computers to the overall corporate objective of the banks, not only in terms of information processing but also on other physical matters.

The financial implication would also need to be considered at this stage.

Generally, strategic planning when concluded would satisfy the following needs:

(i) Amplify the objectives and key tasks to be achieved by the computer-based MIS when introduced.

(ii) Determination of those who would be responsible for the project or random phases of it.

**B    TACTICAL**

This phase commences after the approval of the recommendations agreed during strategic planning phase.

The tactical phase would consider the earlier various resources on short-term basis. It is also at this phase that planning is done on how policy decision could actually be translated into action.

1. **Defining process pattern**: This will include:
   - The organizational structure, the location of the various branches, department and area offices.
   - The existing mode of interaction between personnel department.

2. **Selecting the equipment**: This will include decision on:
   - Micro and mini-computer
   - Mainframe computer
   - The site(s)
   - Software selection

**C    The Operating Plan**

The operating plan covers the day-to-day progress of the project and usually gets time-scale in terms of weeks or months. The aim of operational planning would be to assess the direct effect of the strategic and tactical phases on the available resources.

The plan would concentrate on the focal point of all the earlier plans and ensure that hitches at the operational level are identified and remedial action taken.

In the planning of this phase, consideration must be given to the following:

(a)    Equipment (size, operation, maintenance)

(b)    Personnel (quality, number required, calibre)

(c)    Software (type, source implementation, assuage)

(d)    Infrastructure (space, nature, control)

(e)    Finance (source, disbursement, control)

With the satisfaction of these pre-requisites, it becomes relatively easy to implement the computer-based MIS.

## 2:15    IMPLEMENTATION OF THE COMPUTER-BASE MIS

There are three approaches that could be used, and they are:

  a. The top-down approach

  b. The pilot system approach

c. The programmatic approach

## 2:15:1 THE TOP-DOWN APPROACH

This is the approach is based on the hierarchical nature of the bank and where it has been agreed that core areas of management would be computerized, while at a later stage, other level of operation would be computerized in a decentralized manner. In this implementation, there would be the head office and lastly, the branch management.

**A   HEAD OFFICE OPERATION**

The aim here is to provide the bank's top management with all the information that would quicken and improve on their corporate strategic planning and control. Information may be provided to inform management decision on the following:

(i)   Loans and advances

(ii)   Cash base and treasury management

(iii)   Staff utilization and emoluments

(iv)   Corporate budget appraisal

**B   REGIONAL ADMINISTRATION**

As soon as the computerized MIS has been fully established at the Head Office, the next level would be to install and implement the system in the regions or area offices. In some cases, the information requirements at this level are identical with those produced for the Head Office, the only difference being the scope and areas of coverage. It therefore follows that once the head office operation is satisfactorily concluded, what is required to take off in the regions would be the availability of the resources needed and may be some minor modification to the software. The input to the processing at this level would come from the branches under its supervision and control.

**C      BRANCH MANAGEMENT**

The introduction of computer at this level satisfies different management and operational needs and would therefore be totally different from the information required at the two preceding levels.    However, the following operational needs are appropriate for effective branch management of banks:

   (i)     Customer enquiries

   (ii)    Cheque reference

   (iii)   Balance checking

   (iv)    Turnover assessment.

   (v)     Clearing cheques through clearing house

   (vi)    Processing of periodic payment requests

   (vii)   Savings bank operation

   (viii)  Provision of reports for area offices.

The implementation of a computerized MIS in the branches must be in conjunction with the area offices. In fact, the area offices would identify the branches where implementation could commence.

Generally, one important thing to note in this decentralization of operation is that the interface between information from one level to the other must be properly worked out and agreed to by the steering committee. The modalities of such interface must have formed part of the steering committee's recommendation, which was originally submitted to the board for approval.

## 2:15:2     THE PILOT SYSTEM APPROACH

This is the approach adopted when it must have been decided to adopt universal standards in a centralized processing pattern.

In centralized processing, the first consideration during installation and implementation would be to identify those aspects of banking operations that are central to all levels of control within the bank, and to these activities to be computerized first. After this initial take off, other modules would be added to the same central computer, based on priorities already established by the steering committee.

A possible scheme of installation and implementation would run like this:

1. Take on the primary duty of retail banking in the immediate surrounding of the central computer.
2. Install the central mainframe and the data.
    - Capture equipments at the primary area of operations within the computer's vicinity.
    - General ledger and current account system
    - Saving bank system
3. Add various essential management control report e.g. report on advances and central banks consolidate returns.
4. Include other modules progressively:
    - Periodic payment
    - Transfers
    - Letter of credit and import financing
    - Other remittances
5. Computerize other level of management needs
    - Finance and budget control
    - Treasury management
    - Personnel records

6. Finally, on installation and implementation, whatever method is adopted must have been extensively discussed and appropriate decision taken during the various planning phases.

It then follows that as soon as the approval is granted, the head of the computer department becomes in effect the project leader and ensures that all operational plans are executed to accurately. What this means is that the project leader who in this case is in charge of the functional area of the computer department, must report regularly to the steering committee. This process is continuous until all the banks area activities are computerize.

## 2:15:3  THE PRAGMATIC APPROACH

This is the combination of the two other approaches. It is a hybrid approach usually favoured by the banks.

## 2:16  INTERNAL CONTROL IN A COMPUTERIZED ACCOUNTING SYSTEM –ELECTRONIC DATA PROCESSING (EDP)

With the knowledge of the development and implementation of a computerized MIS, we can now look at internal control in a computer-based information system. Internal control is a sine-qua-non for effective accountability, and with the system being computerized, it stands to pose new challenges to the operations of the computer system

By internal control, we mean not only internal check and internal audit, but the whole system of control, financial and otherwise, established by the management in order to carry on the business of the company in an orderly manner and to safeguard its assets.

According to J. Santock (1978, page 11), internal control is an umbrella beneath which are financial controls, internal check and audit.

Spicer and Pegler (1985, pg 102), believe that where a system is computerized, general controls are needed to ensure proper development and implementation of applications and the integrity of program and data file.

According to Olusanya T (2002),Computer systems have features which differentiate them from non-EDP systems and which, potentially, can create internal control problems. Some of these features and their potential problems are as follows:

(a) Computer systems are more complex than conventional systems.

(b) Computer systems are usually highly integrated. This often results in the concentration of previously separate data processing functions in the EDP department. This development increases the risk of irregularities within the EDP department as the quality of internal check is thereby reduced.

(c) In a computer environment, control procedures which were previously exercised clerically are included in the program logic and are autonomously executed by the computer during processing. The computer will often not leave any documentary evidence of complying with these controls. .

**Internal Controls in an EDP Environment**

It can be seen from the above that while EDP systems offer significance advantages and advancement on data processing, they are fraught with potential dangers. As a result, management should, when setting up an EDP system, carefully consider and establish adequate internal control over the EDP operations. There should be procedures also to ensure that EDP controls are reviewed regularly for effectiveness and efficiency. The broad objectives of the controls will include:

(a) To ensure that EDP operations are carried out in an orderly and efficient manner.

(b) To safeguard the EDP facilities including the central processing, the computer files and the computer peripherals.

(c) To ensure that polices relating to the EDP department are adhered to

**<u>Division Internal Controls in an EDP Environment:</u>**

**Organizational Controls**

These controls relate to the environment within which application and programs are developed, maintained, and they are:

1. **Administrative Controls which can be subdivided into:**

    (a) Segregation of duties

    (b) Control over computer operators

    (c) File controls

    (d) Fire precaution and standby arrangements.

2. **Systems development controls.**

    a. **Hardware.**

    b. **software.**

**Administrative Controls**

a. **Segregation of duties**: This implies that the main function in a transaction is carried out by separate individuals or group of persons. It is aimed at minimizing the risk of errors and fraud inherent in the system. In an EDP environment, the following control relating to the segregation of duties should be established:

- There should be separation of duties between the user department function and the EDP function.

- The main functions within the EDP department should be carried out by separate persons. The functions include:

An organizational chart should be prepared and the principal tasks in the department should be defined and allocated to separate individuals and group of individuals.

The following restrictions should apply within the EDP department whenever practicable:

- Only the control and data preparation section should have access to the documents containing the original data to be processed by the computer.
- Access to the computer room should be restricted to authorized person at authorized times.

The extent to which the above segregation of duties can be achieved in practice will vary with the size of staff in the EDP department. Small businesses are generally constrained by inadequate human and financial resources. Under such circumstances, not all the jobs described above can be carried out by different groups of people at all times. In these cases the effect of inadequate segregation of duties should be balanced by increased supervision by responsible officials of the business.

    **b.   Control Over Computer Operators:**

The computer operators are responsible for operating the equipment. This should be in accordance with operating rules of the enterprise. The work of computer operators includes:

(a) Setting up the equipment for each run (i.e. loading the input devices, setting up the magnetic storage devices and loading the correct stationery into the printer);

(b) Running programs on the computer system;

**The objectives of control over operators are:**

- To ensure that the computer is used for bona fide businesses only.
- To ensure that the computer is used effectively and efficiently
- To ensure that transactions are processed on timely basis.
- To minimize the risk of operators' errors and fraud.
- operating instructions for each program;

  c. **File Control**

Modern computer files are maintained in a magnetic backing storage device such as magnetic tapes and disks. In view of the large number of records and programs held on magnetic backing storage devices in most EDP systems, it is necessary to provide adequate controls to ensure that:

- Only the correct file can be used for processing
- No file can be used for an unauthorized purpose

- That files are not unintentionally overwritten
- That files can be reconstructed if they are lost or their content are corrupt.

**Controls that may be established to achieve the above objectives include:**

- Maintaining computer files and programs in a secure manner, for example, by the use of locks and keys.
- Maintaining computer files and programs under a conducive atmospheric condition. For example, the file library should be properly air-conditioned, equipped with smoke and dust detectors, humidity and temperature recorders and fire fighting facilities.

## 2.17   COMPUTER FRAUD IN BANKS

The computer system, being an electronic device, though very intelligent, relies upon data input and stored in it to produce relevant and incisive information. Hence, the slogan "Garbage in, Garbage out". In essence, the computer is not in a position to detect whether the data it receives is fraudulent or not. As a result, dubious staff of banks capitalize on this weakness of the computer system to defraud their organization and cause them to lose a lot of money.

Another disadvantage is that, once an unauthorized access is gained to the database all that is required is a single command to debit all account or a particular account, and fraudulently withdraw the money. To preclude the occurrence of fraud, banks will need to set up a high standard internal control mechanism, which will be proficient in detecting and preventing fraud.

**MEASURES TO ADOPT IN CHECKMATING COMPUTER IN BANKS**

In this aspect, no one person should be in charge of a particular process. That is, where one person's duty ends, the other's duty begins. This menace can be curtailed if the management strives:

a) To ensure that the computer is used for bona fide businesses only.

b) To ensure that transactions are processed on a timely basis.

c) To ensure that works to be carried out on a computer is scheduled.

d) To ensure that the duties of the operations should be rotated.

e) To ensure that there is a minimum of two computer operators per shift.

**2.18    THE PROBLEM OF OBSOLESCNCE OF SYSTEMS**

This problem could be seen from the history of the united bank for Africa plc, which,over its many years of experience, has changed from centralized system prior to 1970 to Branch accounting system in the mid 1970s, with its concentrations of information system at it head office, from where printout and the information output are produced and distributed to its various branches in a centralized manner. Subsequently, UBA Plc started to decentralize its operations on a stand-alone basis in 1986 .

In 1992, the bank started full computerization of its branch network, making each branch to be fully on-line and independent of both the Head Office and other branches, in the generation and processing of data, based on "Branch Accounting Information System (BRAINS) software. The bank has also made inter-branch communication very easy through its "very small Aperture Terminal" (VSAT) satellite. This is the means through which customers can transact businesses with branches other than that where they maintain their account.

Of available system that will meet and satisfy their needs before doing so.

## 2.19. THE LITERATURE ON COMPUTERIZATION

According to Rob Kling (1996), "when specialists discuss computerization and work, they often appeal to a strong implicit image about the transformation of work in the last one hundred years, and the role that technologies have played in some of those changes. In the nineteenth-century North America, there was a major shift from farms to the offices as the primary workplaces. Those shifts often associated with the industrial revolution-continued well into the early twentieth century. Industrial technologies such as the steam engine played a key role in the rise of industrialization.

Our twentieth-century economy has been marked by the rise of human service jobs, in area such as banking, insurance, travel, education, and health. And many of the earliest commercial computer system were bought by large service organization such as banks and insurance companies. By some estimates, the finance industries bought about 30% of the computer hardware in the United States in 1980s. During the last three decades, computer use has spread to virtually every kind of workplace, although large firms are still the dominant investors in computer-based system. Since offices are the predominant site of computerization, it is helpful to focus on offices in examining the role that these systems play in altering work.

In the early twentieth century, the technologies and organization of office work underwent substantial change. Firms began to adopt telephones and typewriters, both of which had been recently invented. By the 1930s and 1940s, many manufacturers devised electromechanical machines to help manipulate, sort, and tally specialized paper records automatically. Some of the more expensive pieces of equipment, such as specialized card-accounting machines, were much more affordable and justifiable in organizations that centralized their key office activities.

During the 1980s and early 1990s, virtually every organization bought PCs and workstations. But "the PC revolution" did not merely change the nature of office equipments- it expanded the range of people who use computers routinely to

# CHAPTER THREE

## 3.1 SELECTING THE SAMPLE

This research project is aimed at making a definite statement on the effect of computerization on the Nigerian banking industry. And, in an attempt to do this effectively, the researcher zeroed in on United Bank for Africa Plc as a sample, which he believes will be representative of the Nigerian banking industry. The sample was selected bearing in mind how unwieldy it will be to study the entire Nigerian banking industry, and the need to come out with a reliable research finding.

## 3.3 RESEARCH DESIGN

The researcher made use of field study or survey approach. This was based on the use of personal interview and administration of questionnaire. The questionnaire was designed in such a way that it contained a blend of fixed alternative questions which were meant to limit the responses of respondents to stated alternative and open- ended questions which offered the respondents opportunities of expressing their opinions.

The questionnaire consists of three parts. The first part contained background information, compromising questions on general operation of computer department and it acceptability to the entire staff and management. The second part comprised questions which helped to find out the application and usage of computer including types and availability of hardware and software.

The third part generally tries to find out the positive impact of computer application on the selected bank and short-comings of computerization as well. Personal interview were used mainly where the questions are of such technical nature that only a specialist could deal with. This helped the researcher in his search for objectivity and increased validity of his findings and conclusion.

The researcher considered a judicious mix of personal interview, free response questions and fixed alternative questions more desirable in producing a thorough description of the other processes occurring in the organization.

## 3.3   SAMPLE DESIGN

In selecting those elements from which information was obtained, the researcher used the stratified random and simple random sampling method. By stratified random sampling technique, the research divided the organization, which is the population under study, into four strata based on a simple random criterion. Afterward, a simple random sample was taken from each stratum and sub-samples are joined to form the total sample for the study. Considering the nature of this study, the researcher studied forty respondent randomly selected from the strata (organization) but only thirty-four were able to respond.

## 3.4 COLLECTION AND CHECKING OF DATA

The researcher collected the necessary data by personally administering the questionnaire prepare for this purpose, on the respondent. The researcher also interviewed some members of staff of the bank (mostly top and technical officers) on some important issues. In all, the researcher administered forty questionnaires to the bank and respondents were given two working days to read and fill them. At the end of the period, the researcher went personally to collect the completed questionnaire and was only able to collect thirty four. The rest of the questionnaire couldn't be collected since they were not completed.

Therefore, the researcher tested the completed questionnaire for validity and reliability.

## 3.5 METHODS OF DATA ANALYSIS

The procedures for data analysis utilized by the researcher include the following:

(1). Placing each items in its appropriate category reflective of the hypothesis to be tested.

(2). Tabulation of data.

(3). Performing Statistical Computation: The statistical computations used by the researcher included percentage analysis and empirical test of general characteristics of the respondents.

Also, the responses to key questions and statements in the questionnaire will be analyzed using percentage (%). The purpose of this is to simplify the problem of comparisons.

In the empirical test of hypothesis, the researcher will use the chi-square technique. In applying the chi-square technique, the hypothesis to be tested will first and foremost be stated as a null hypothesis (Ho), and an alternative hypothesis (H1) will follow suit.

Thereafter, the value of the chi-square will be calculated using the formula:

$$X^2 = \frac{(O-E)^2}{E}$$

Where:

O = Observed frequency

E = Expected frequency

O-E = Absolute value of the difference between the frequencies (deviation)

$(O-E)^2$ = Deviation squared.

$\frac{(O-E)^2}{E}$ = Sum of all deviation squared and weighted.

By this test, if the computed value of $X^2$ is larger than the tabulated value of $X^2$, we reject the null hypothesis (Ho), and accept the alternative hypothesis (H1). In order to apply the $X^2$ test, the degrees of freedom and the level of significant are very important. The degree of freedom refers to the component of chi-square ($X^2$) which are free to vary at random and independently. Once the border total has been specified, the degree of freedom is ascertained by the formula:

M= r-1

Where r = class of observations in the particular table of our concern. The researcher will utilize 0.05 and 0.01 level of significance in testing the hypothesis. Finally, the calculated $X^2$ and the tabulated $X^2$ will be compared and a decision taken.

**DECISION RULE:** $X^2$

Accept Ho if tabulated $X^2$ is greater than calculated $X^2$; reject H1.
Reject Ho if tabulated $X^2$ is less than calculated $X^2$; accept H1.

# CHAPTER FOUR

This chapter has to do with the classification of the ways in which data obtained via the questionnaire were ordered and analyzed so as to obtain the desired result.

## 4.1 ANALYSES OF GENERAL CHARACTERISTICS OF RESPONDENTS.

**Table 4.1**

| DISTRIBUTION | NUMBER | PERCENTAGE (%) |
|---|---|---|
| Questionnaire Returned | 34 | 85 |
| Questionnaire not Returned | 6 | 15 |
| **TOTAL** | **40** | **100** |

Source: field survey (2011)

From the table 4.1 above, a total of 40 questionnaires were administered. A total of 34 which represents 85% of the total questionnaires administered were returned. Only 6 which represent 15% of the questionnaires were not returned.

## 4.2 DEPARTMENT OF RESPONDENTS
**Table 4.2**

| DEPARTMENT | NUMBER | PERCENTAGE (%) |
|---|---|---|
| Computer | 15 | 44.10 |
| Account | 10 | 29.43 |
| Marketing | 9 | 26.47 |
| TOTAL | 34 | 100 |

Source: field survey (2011)

From the table 4.2 above, 15 respondents representing 44.10% work in the computer department of the bank. While 10 respondents representing 29.43% work in the Account department. And, 9 of the respondents representing 26.47% work in the Marketing department.

## 4.3 EVALUATION OF WHETHER THE BANK IS COMPUTERIZED

**Table 4.3**

| OPTION | NUMBER | PERCENTAGE |
|---|---|---|
| YES | 34 | 100 |
| NO | 0 | 0 |
| **TOTAL** | **34** | **100** |

Source: field survey (2011)

From the table 4.3 above, all the respondents (100%) agreed that the bank is computerized. Therefore, the bank is highly computerized.

## 4.4 EVALUATION OFWHETHER COMPUTERIZATION HAS BROUGHT EFFICIENT SERVICE DELIVERY IN THE BANK.

**Table 4.4**

| OPTION | NUMBER | PERCENTAGE |
|---|---|---|
| YES | 33 | 97.06 |
| NO | 1 | 2.94 |
| **TOTAL** | **34** | **100** |

Source: field survey (2011).

From the table 4.4 above, 33 respondents (97.06%) agreed that computerization has brought about efficient service delivery in the bank. While only 1 respondent representing 2.94% disagreed that computerization has brought efficient service delivery in the bank. This implies that computerization has brought efficient service delivery in the bank.

## 4.5 EVALUATION OF WHETHER COMPUTERIZATION CAN LEAD TO JOB-LOSS IN THE BANK

**Table 4.5**

| OPTION | NUMBER | PERCENTAGE (%) |
|---|---|---|
| YES | 4 | 11.76 |
| NO | 30 | 88.24 |
| **TOTAL** | **34** | **100** |

Source: field survey (2011)

From the table 4.5 above, 4 respondents, representing 11.76% of the respondents agreed that computerization can lead to job loss in the bank. While 30 of the respondents (88.24%) disagreed that computerization can lead to job loss in the bank. This means that computerization cannot lead to job loss in the bank.

## 4.6 EVALUATION OF WHETHER COMPUTERIZATION HAS INCREASED THE INCIDENCE OF FRAUD IN THE BANK.

**Table 4.6**

| OPTION | NUMBER | PERCENTAGE |
|---|---|---|
| YES | 2 | 5.88 |
| NO | 32 | 9.12 |
| TOTAL | 34 | 100 |

Source: field survey (2011)

From the table 4.6 above, 2 of the respondents representing 5.88% agreed that computerization has increased the incidence of fraud in the bank. While 32 of the respondents representing 94.12% disagreed that computerization has increased the incidence of fraud in the bank. Therefore, computerization has not increased the incidence of fraud in the bank.

## 4.7 EVALUATION OF THE EFFECT OF COMPUTERIZATION ON THE BANK'S INTERNAL CONTROL SYSTEM.

**Table 4.7**

| EFFECT | NUMBER | PERCENTAGE (%) |
|---|---|---|
| Positive | 33 | 97.06 |
| Negative | 1 | 2.94 |
| TOTAL | 34 | 100 |

Source: field survey (2011)

From the table 4.7 above, 33 respondents representing 97.06% agreed that computerization has had a positive effect on the bank's internal control system. While 1 respondent representing 2.94% of the respondents was of the opinion that computerization has had a negative effect on the bank's internal control system. This means that computerization has a positive impact on the bank's internal control system.

## 4.8 EVALUATION OF WHETHER THE STAFFING OF THE BANK'S COMPUTER DEPARTMENT IS ADEQUATE

**Table 4.8**

| OPTION | NUMBER | PERCENTAGE (%) |
|---|---|---|
| YES | 31 | 91.18 |
| NO | 3 | 8.82 |
| TOTAL | 34 | 100 |

Source: field survey (2011)

From the table 4.8 above, 31 respondents representing 91.18% of the respondents agreed that the staffing of the banks computer department is adequate. While 3 respondents (8.82%) disagreed that the staffing of the banks computer department is adequate. Therefore, the staffing of the bank's computer department is adequate.

## 4.9 EVALUATION OF WHETHER THE COMPUTER HARDWARE AND SOFTWARE OF THE BANK IS RELIABLE.

**Table 4.9**

| OPTION | NUMBER | PERCENTAGE (%) |
|---|---|---|
| YES | 34 | 100 |
| NO | 0 | 0 |
| TOTAL | 34 | 100 |

Source: field survey (2011)

From the table 4.9 above, all the respondents (100%) agreed that the hardware and software employed by the bank are reliable. This implies that the computer hardware and software of the bank is reliable.

## 4.10 EVALUATION OF WETHER COMPUTERIZATION SHOULD BE ENCOURAGED IN THE BANK.

**Table 4.10**

| OPTION | NUMBER | PERCENTAGE (%) |
|---|---|---|
| YES | 33 | 97.06 |
| NO | 1 | 2.94 |
| **TOTAL** | **34** | **100** |

Source: field survey (2011)

From the table 4.10 above, 33 respondents, representing 97.06% of the respondents, agreed that computerization should be encouraged. While only 1 respondent (2.94%) disagreed that computerization should be encouraged.

## TEST OF HYPOTHESIS

This will be done using table 4.4.

$H_o$: Computerization has not improved service delivery in the United Bank for Africa Plc.

$H_1$: Computerization has improved service delivery in the United Bank for Africa Plc.

## Table 4.4

Evaluation of whether computerization has brought efficient service delivery in the bank.

| Option | Number | Total |
|---|---|---|
| Yes | 33 | 33 |
| No | 1 | 1 |
| **TOTAL** | **34** | **34** |

Source: field survey (2011)

$E_i$ = Expected frequency

$O_i$ = Observed Frequency

$$E_i = \frac{31+1}{2} = \frac{34}{2} = 17$$

| Cell | Oi | Ei | Oi-Ei | (Oi-Ei)² | (Oi-Ei)²/Ei |
|---|---|---|---|---|---|
| C$_{l1}$ | 33 | 17 | 16 | 256 | 15.06 |
| C$_{12}$ | 1 | 17 | -16 | 256 | 15.06 |
| E | 34 | | | | 30.12 |

$$X^2 = \frac{E(O_i - E_i)^2}{E_i} = 30.12$$

$X^2$ Tabulated:
Degree of freedom = n-1 = 2-1=1. The table of value of $X^2$ at 5% level of significance and degree of freedom is 3.841.

**THE DECISION RULE**

If the computed $X^2$ is less than the table value of $X^2$, we accept the null hypothesis (Ho), otherwise reject it.

**CONCLUSION:**

Since the computed $X^2$ is higher than the table value of $X^2$, we reject the null hypothesis (Ho), but accept the alternative hypothesis (Hi). Here, it can be concluded that computerization has brought efficient service delivery in the United Bank for Africa Plc, which is possibly applicable to all other banks in the industry

# CHAPTER FIVE

# RESEARCH FINDINGS, RECOMMENDATION AND CONCLUSION

## 5.0 INTRODUCTION

The research on the effect of computerization on the Nigerian banking industry was carried out by the researcher with the aim of making informed and encyclopaedic reference on the effect of computerization on the Nigerian banking industry, bringing out succinctly its pros and cons, and to find out what could be done to gain the utmost advantage from this wonderful electronic device which has changed the face of banking in Nigeria and beyond.

In the process of this research, a lot of people were interviewed, questionnaires were distributed to both programmers and computer users and the following is the synopsis of the finding of the study.

## 5.1 RESEARCH FINDING

The research was focused on the effect of computerization on the Nigerian banking industry, with particular interest in the United Bank for Africa Plc, which is considered representative of the happenings in the industry.

In order to make an informed inference, on the study, various relevant literatures were consulted and field study carried out to obtain relevant data. The data collected were analyzed and various hypotheses formulated were subjected to test and verification. The study revealed the following:

1. United Bank for Africa Plc is fully computerized.

2. Computerization fully and effective supports the operation of the United Bank for Africa Plc, and enables the bank to meet its customers' demand.

3. The study revealed that a good percentage of the staff of United Bank for Africa Plc is computer literate, while others need further training.

4. The staff appreciates computerization and see computer as a device meant to assist them in carrying out their job effectively, as against seeing it as a replacement for manpower.

5. Computerization is fully embraced by the entire staff and management of the bank.

6. The study revealed that the bank uses on-line and real-time instead of batch processing.

7. The computer hardware and software being used by the bank are highly reliable.

8. The study revealed that the computer department of the bank requires further staffing.

9. The level of training given to computer users and operators is just adequate with equal room for further improvement.

10. The respondents were fully of the opinion that computerization has greatly improved service delivery in the bank.

11. With computerization, the services of the bank have been quite timely, efficient and effective.

12. The study revealed that internal control has not been negatively affected by computerization.

13. The study revealed that computerization has greatly reduced the incidence of fraud in the bank.

14. The respondents want computerization encouraged in the Nigerian banking industry.

15. The study revealed that the major set-back of computerization is the heavy cost implication of the process, and the need for adequate and continuous training of staff which is quite expensive.

16. Computerization has reduced the transaction-processing time in the Nigerian banking industry.

## 5.2   RECOMMENDATIONS

Having carried out the research on the effect of computerization on the Nigerian banking industry diligently, the researcher hereby recommends that:

i) Banks should embrace computerization fully, considering the immense benefits that accrue from its use.

ii) Banks should study extensively the systems available before choosing the one to adopt in order to avoid the huge cost implication of changing from one computerization system to another.

iii) Banks should map out a good training program for their staff to help them keep abreast of developments in computer technology, as it relates to banking operations

iv) Banks should ensure availability of trained personnel to man their computer departments.

v) Banks should ensure adequate back-up procedures as regards the availability of computer spare parts, hardware and software.

## 5.3 CONCLUSION

The introduction of computers into the Nigerian banking industry, though spanning over a relatively short period of time have had a great and beneficial impact on the industry. Transactions now take very short time to be processed, customers' expectations are being met, and volume of business transaction has increased tremendously, both for the banks and its customers.

Considering the above positive scenario, the researcher is of the opinion that computerization should be encouraged and advises that banks go the computer way. Government should monitor development in the industry to avoid dumping of outdated computers in the banking industry.

Furthermore, the researcher is of the opinion that further researcher work should be encouraged in this direction using the work already done as a basis and a guide.